\def\k{{\bf k}}
\def\be{\begin{equation}}
\def\ee{\end{equation}}
\def\ber{\begin{eqnarray}}
\def\eer{\end{eqnarray}}
\def\bers{\begin{eqnarray*}}
\def\eers{\end{eqnarray*}}

\def\JPC{J. Phys. C}
\def\JPCM{J. Phys.: Condens. Matter}
\def\PR{{ Phys. Rev.}\ }
\def\PRL{{ Phys. Rev. Lett.}\ }
\def\JPC{{ J. Phys. C: Solid State Phys}\ }
\def\JPCM{{ J. Phys.: Condens. Matter}\  }

\documentclass[aps,prb,showpacs,floatfix,a4paper,twocolumn]{revtex4}
\usepackage{amsmath}
\usepackage{amsfonts}
\usepackage{amssymb}
\usepackage{subfigure}
\usepackage{graphicx}
\begin{document}

%%%%%%%%%%%%%%%%%%%%%%%%%%%%%%%%%%%%%%%%%%%%%%%%%
%               TITLE 
%%%%%%%%%%%%%%%%%%%%%%%%%%%%%%%%%%%%%%%%%%%%%%%%%
\title[Effect of disorder on the electronic properties of graphene: A theoretical approach]
{Effect of disorder on the electronic properties of graphene: A theoretical  approach}

\author{Aftab Alam}
\email[emails: ]{aftab@ameslab.gov}
\affiliation{Division of Materials Science and Engineering, Ames Laboratory, Ames, Iowa 50011, USA}
\author{Biplab Sanyal}
\affiliation{Department of Physics and Astronomy, Uppsala University, Box-516, 75120 Uppsala, Sweden}
\author{Abhijit Mookerjee}
\affiliation{Department of Condensed Matter Physics and Materials Sciences, S.N.Bose National Centre for Basic
Science, JD-III, Salt Lake City, Kolkata 700098, India}

%%%%%%%%%%%%%%%%%%%%%%%%%%%%%%%%%%%%%%%%%%%%%%%%%
%               ABSTRACT 
%%%%%%%%%%%%%%%%%%%%%%%%%%%%%%%%%%%%%%%%%%%%%%%%%
\begin{abstract} 
In order to manipulate the properties of graphene, its very important to understand the electronic structure in presence of disorder. We investigate, within a tight-binding description, the effects of disorder in the on-site (diagonal disorder) term in 
the Hamiltonian as well as in the hopping  integral (off-diagonal disorder) on 
the  electronic dispersion and density of states by the augmented space 
recursion method. Extrinsic off-diagonal disorder is shown to have dramatic 
effects on the two-dimensional Dirac-cone, including asymmetries in the band 
structures as well as the presence of discontinuous bands (because of resonances) in certain limits. Disorder-induced broadening, related to the scattering 
length (or life-time) of Bloch electrons, is  modified significantly with  increasing strength of disorder. We propose that our methodology is suitable for
the  study of the effects of disorder in other 2D materials, such as a boron nitride mono layer.
\end{abstract}
\date{\today}
\maketitle

%%%%%%%%%%%%%%%%%%%%%%%%%%%%%%%%%%%%%%%%%%%%%%%%%
%                INTRODUCTION
%%%%%%%%%%%%%%%%%%%%%%%%%%%%%%%%%%%%%%%%%%%%%%%%%
\section{Introduction}
{\par}  Graphene, a two-dimensional allotrope of carbon, plays a central role in 
providing a basis for understanding the electronic properties of other carbon
allotropes. Being one of the thinnest and the strongest material 
ever measured, graphene has attracted the attention of the materials research 
community\cite{Castro09}  in the recent past.
One of the most interesting aspects of graphene is that
its low energy dispersion closely resembles  the Dirac spectrum of massless 
fermions. This particular type of dispersion provides a bridge between condensed
matter physics and quantum electrodynamics (QED) for massless fermions. Of course
 in graphene, the Dirac fermions move with a much smaller speed.

 Because of its unusual electronic and structural flexibility, properties of 
graphene can be controlled chemically or structurally in many different
ways. For example, deposition of metal atoms\cite{Calandra07} on top of the graphene sheet,
incorporating  
other elements like boron and nitrogen\cite{Martins07} randomly in the parent 
structure, either interstitially or substitutionally and using different substrates.\cite{Calizo07}
  Because disorder is unavoidable in any material, there has
been an increasing interest in understanding how disorder affects the physics of 
electrons in graphene.\cite{uppsala} Disordered graphene based derivatives can probably be 
referred to as functionalized graphene suitable for specific applications. 
``{\it Graphene paper}"\cite{Dikin07} is a spectacular example of how 
important such functionalization could be. 

There can be many different sources of disorder in graphene
including both intrinsic as well as extrinsic. 
Intrinsic sources may include surface ripples and topological defects.
Extrinsic disorder comes in the  form of  vacancies, adatoms, quenched
substitutional atoms and extended defects such as edges and cracks. Another
way of introducing disorder is by ion-irradiation that produces
complex defect structures in the graphene lattice.\cite{arkady} Graphene in an
amorphous form may increase the metallicity too.\cite{erik}

To have a theoretical description of graphene's electronic structure, 
one may begin with the Kohn-Sham equation and a tight-binding
representation whose basis is labeled by the sites of the underlying Bravais
lattice. Disorder may enter the matrix representation of the Hamiltonian in
two ways : vacancies, dopants and adatoms predominantly  cause a random 
change in the local single-site energy (disorder in the diagonal terms) but 
through the overlap such defects 
modify the hopping integrals between different sites (disorder in the 
off-diagonal terms)
causing an effective random change in the distance or angle between the
 bonding orbitals. Thus diagonal and off-diagonal  disorders simultaneously
occur and are correlated. Model calculations which take them to be independent 
are qualitatively in error. As far as diagonal disorder is concerned, it acts as a simple chemical
potential shift of the Dirac fermion i.e. shifts the Dirac point locally.
Theoretical study of such disorder is rather simple and has indeed received 
attention and success, reported in literature.\cite{uppsala}$^-$\cite{35}  A proper inclusion of
  off-diagonal disorder, on the other hand, is non-trivial and requires more sophisticated
approaches.

\begin{figure}[t]
\centering
\includegraphics[scale=0.28]{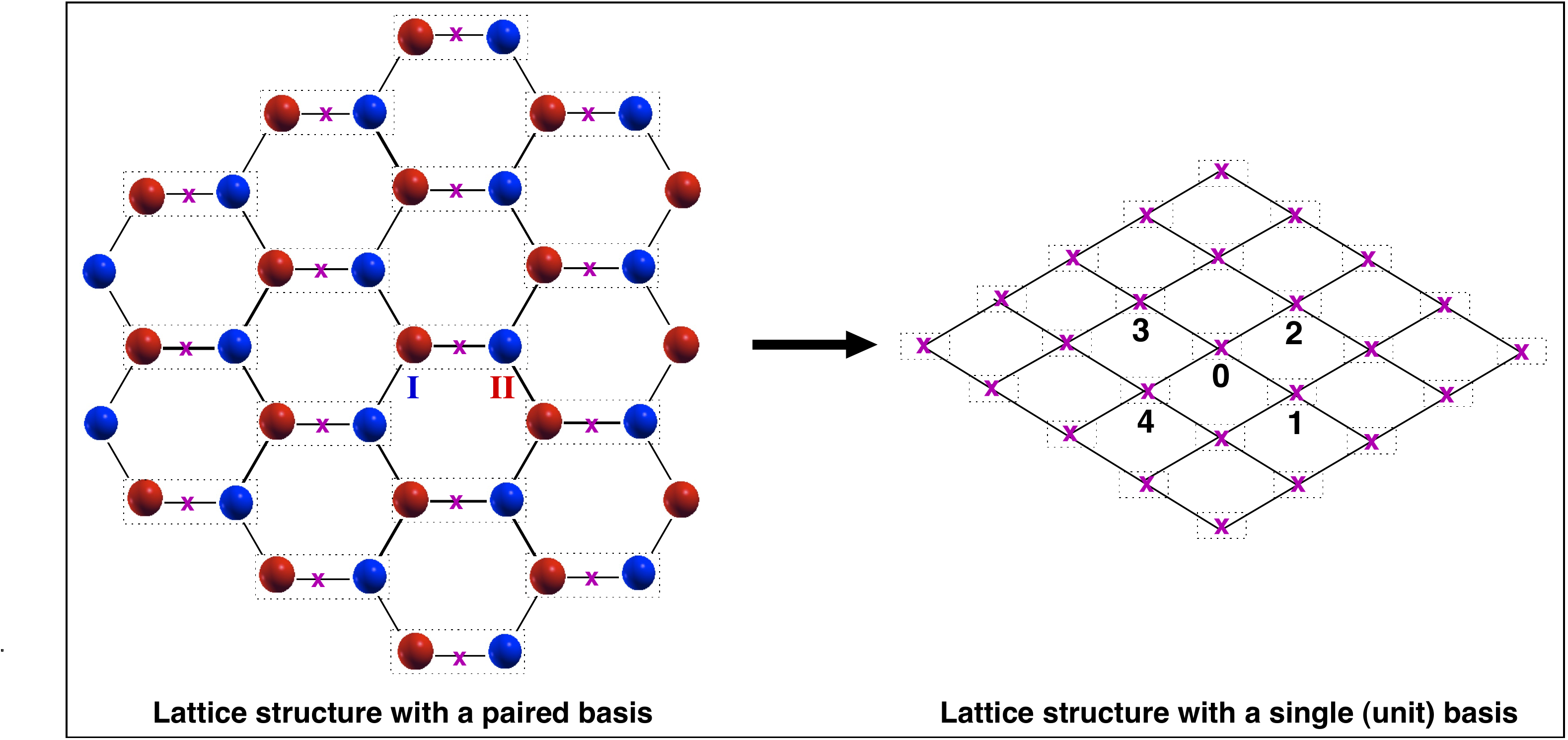}
\caption {(Color online) (Left) The standard honeycomb lattice with a basis of two atoms per unit cell. (Right) The underlying  rhombic Bravais lattice which
 becomes the honeycomb lattice when a pair of atoms decorate each site.}
\label{fig1}
\end{figure}

\begin{figure}[b]
\vskip 0.5cm
\centering
\includegraphics[scale=0.3]{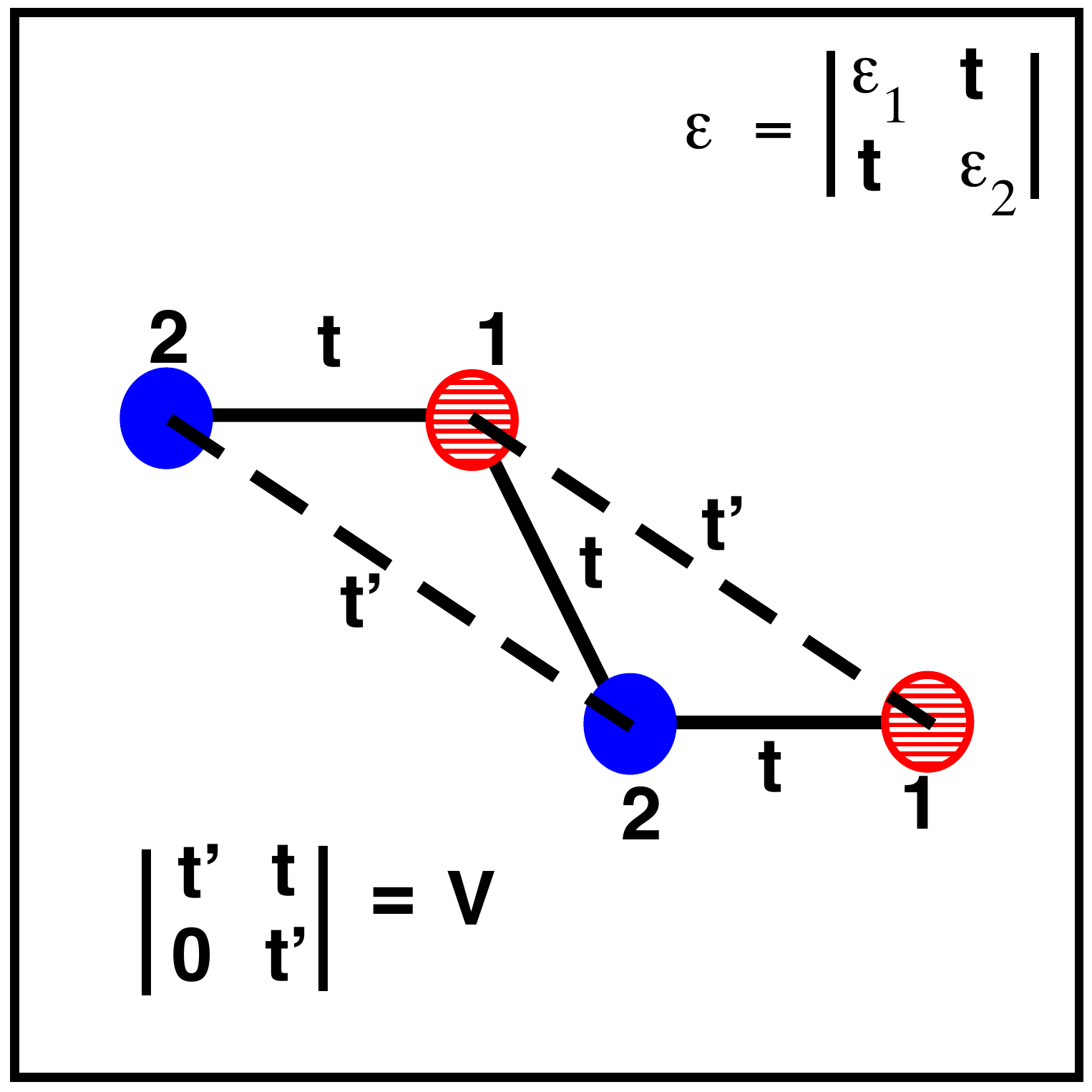}
\caption {(Color Online) Nearest neighbor overlaps on the rhombic lattice}
\label{fig2}
\end{figure}

{\par}
Till date there have been numerous attempts at studying the effects of disorder in
graphene.\cite{uppsala}$^-$\cite{35} Among others, the methods used to study effects of disorder included the averaged t-matrix approximation (ATA)\cite{35} and the 
coherent potential approximation(CPA).\cite{CPA} The first one is not self-consistent and hence inaccurate. The latter is a single-site mean field approximation with all
its attendant problems.\cite{CPA} Several others have used exact diagonalization of huge clusters and
 the real-space recursion of Haydock {\emph {et al.}}\cite{35,hhk} Both these techniques actually calculate the 
density of states (DOS) for  specific configurations of the system followed by  direct averaging over a large number of 
configurations. Since each of the configurations has periodic boundary conditions, the averaged spectral
 function is always a collection of delta functions  and the disorder induced life-time effects cannot be probed. The recursion on the lattice probes mainly the real-space effects of disorder.

From the theoretical perspective, dealing with disorder
has had a long history. As mentioned earlier, one of the most successful and frequently  used approaches is
the single-site, mean field CPA.\cite{CPA}  
However, as the name itself suggests, it
is a single site approximation and cannot adequately take into account the
effects of correlated configuration fluctuations. In particular the CPA is 
inaccurate at low dimensions. In one dimension it is shown to be inadequate by Dean\cite{dean} some time ago. Among the hierarchy of the
generalizations of CPA, only a few approaches have maintained the  necessary Herglotz
analytic properties and lattice translation symmetry of the configuration 
averaged Green's function. These  include the non-local CPA,\cite{NLCPA}
 the special quasi-random structures (SQS),\cite{SQS} the 
locally self-consistent multiple scattering approach (LSMS),\cite{LSMS} and
the three methods based on the 
augmented space formalism proposed by one of us \cite{Mookerjee73} :
 the traveling cluster
approximation (TCA),\cite{TCA} the itinerant coherent potential approximation 
(ICPA),\cite{ICPA} and the augmented space recursion (ASR).\cite{ASR}
Over the years ASR has proved to be  one of the most powerful techniques,
 which can accurately 
take into account the effects of correlated fluctuations arising out of the disorder in the 
local environment. This is reflected in a series of studies in the past
e.g. the effects of local lattice distortion as in CuBe,\cite{Latt_dist} 
 short-range ordering due to local chemistry,\cite{SRO} the phonon problem
\cite{phonon} with essential off-diagonal disorder in the dynamical matrices,
and electrical and thermal transport properties\cite{transport} in disordered
alloys. 

In this communication, we present  a 
theoretical tight-binding model to study the effects of disorder in graphene.
 Disorders studied were mainly of two forms : substitutional disorder
\cite{10,32}$^-$\cite{34} and vacancies.\cite{4,35}
Unlike earlier models, both the diagonal and off-diagonal disorders are included
on the same footing. 
 The present formalism is based on the augmented space
 recursion.\cite{ASR}  Although recursion has been 
used to study graphene before,   we want to emphasize that in {\sl all}
those applications recursion was carried out on a Hilbert space ${\cal H}$ spanned by the tight-binding basis representing the Hamiltonian. In
augmented space recursion, we recurse in the space of all possible configurations which the Hamiltonian may assume in the disordered system. For a homogeneously 
disordered binary alloy, this configuration space is isomorphic to that of a
spin-half Ising model. The augmented space theorem \cite{Mookerjee73} then connects configuration averages to a specific matrix element in that space of configurations. 

The novel approach in this work is that we shall make use of 
the translation symmetries in augmented space (for homogeneous disorder) to carry out  recursion in reciprocal space. This will directly give us the spectral function from  which we extract the 'fuzzy' band structure. The inclusion of 
 the effects of configuration fluctuations of the immediate environment gives us self-energies
which are strongly $\k$ dependent, unlike the CPA. In order to make a systematic study, we present results for 
combinations of both strong and weak  diagonal and off-diagonal 
disorder. The combined effects show dramatic changes in the location and
topology of the Dirac-like dispersion and the DOS. Special
emphasis has been  given to the non-trivial inclusion of off-diagonal disorder,
in which case the averaged Bloch spectral function comes out to be significantly 
broadened, multiply peaked, and asymmetric in certain limits where the presence of 
resonances leads to 
 discontinuous dispersion. The interesting interplay of the two kinds
of disorder on full-widths at half maxima (FWHM) (related to the life-time
of Bloch electrons in a disordered system) is also shown.

{\par} The rest of the paper is organized as follows. In Sec. II, we introduce 
the basic formalism. Sec. III is devoted to results and discussions. 
Concluding remarks are present in Sec. IV.

\section{Formalism}
The most general tight-binding Hamiltonian for electrons in graphene can be represented as,
\begin{equation}
H=\sum_{R\alpha_s}\sum_{R'\alpha_{s'}}\left\{ \epsilon_{R}^{\alpha_s}\delta_{RR'}\delta_{ss'} P_{R}^{\alpha_s} +  V_{RR'}^{\alpha_{s}\alpha_{s'}} T_{RR'}^{\alpha_{s}\alpha_{s'}}\rule{0mm}{4mm}\right\},
\label{eq1}
\end{equation}
where $R, R'$ denotes the position of the unit cell of the lattice, $\alpha_s$
denotes the $\alpha$-th atom on the $s$-th sublattice. The actual atomic
position is $R+\zeta^{\alpha_{s}}$, where $\zeta^{\alpha_{s}}$ is the position 
of the $\alpha$-th atom on the $s$-th sublattice. $\epsilon_{R}^{\alpha_s}$ is
the on-site energy describing the scattering properties of the atomic potential
at $R+\zeta^{\alpha_{s}}$, and $V_{RR'}^{\alpha_{s}\alpha_{s'}}$ is the hopping
integral between  $R+\zeta^{\alpha}$ and $R'+\zeta^{\alpha'}$. 
$P$ and $T$ are the projection and transfer operators in the 
Hilbert space spanned by the tight-binding basis $\vert R\alpha_s\rangle$.

{\par} The above Hamiltonian $H$ describes electrons in the original honeycomb
lattice of ion-cores, as shown in the left panel of Fig. \ref{fig1}. The two 
inequivalent sublattices (shown by red and blue spheres) are distinguished 
from each other. The underlying Bravais lattice is the rhombic lattice shown
in the right panel of Fig. \ref{fig1}. Looking at Fig. \ref{fig2} we can 
simplify Eqn.(\ref{eq1}) further and write the Hamiltonian elements as $2\times 2$ matrices :

\begin{equation}
H=\sum_{R} \underline{\underline{\varepsilon}}_{R}\ P_{R} + \sum_{R\ne R'} \underline{\underline{V}}_{RR'} T_{RR'},
\label{eq2}
\end{equation}
where $\underline{\varepsilon}_{R}$ and $\underline{\underline{V}}_{RR'}$, instead of being
scalar for a single-band problem, are now $2\times 2$ matrices given by,
\begin{eqnarray}
\underline{\underline{\varepsilon}}_{R} = \left(\begin{array}{cc}
\varepsilon_{1} & t \\
t          & \varepsilon_{2}   \end{array}\right) \ 
\underline{\underline{V}}_{01} = \underline{\underline{V}}_{02} = \left(\begin{array}{cc}
t' & 0 \\
t  & t' \end{array}\right)\nonumber\\
\underline{\underline{V}}_{03} = \underline{\underline{V}}_{04} = \left(\begin{array}{cc}
t' & t \\
0  & t' \end{array}\right),
\label{eq3} 
\end{eqnarray}
where $\varepsilon_{1}$ and $\varepsilon_2$  are the on-site energy on the two 
sublattices, $t$ and $t'$ are the nearest neighbor and the next nearest neighbor
 hopping energies. $\underline{\underline{V}}_{0I}$ are the hopping matrices between the 
central site $0$ and its four neighboring sites $I$ (in the rhombic lattice) 
 as shown in the right panel of 
Fig. \ref{fig1}. Because the next nearest neighbor hopping $t'$ is usually 
very small compared to $t$, we shall treat the disorder effects only in the 
nearest neighbors.

For a system with substitutional disorder, the most general statement we can make is 
that the occupation of the lattice sites in each inequivalent sublattice can be
 different. For binary disorder in both the sublattices, we may  introduce two 
random occupation variables $n_{R}^{I}$ and $n_{R}^{II}$ associated with the 
sublattices $I$ and $II$ such that,
\[ n_{R}^{I} = \left\{ \begin{array}{ll}
1 & \mbox{if $R \in A$ with probability x$_A$ }\\
0 & \mbox{if $R \in B$ with probability x$_B$ }
\end{array} \right. \]
and
\[ n_{R}^{II} = \left\{ \begin{array}{ll}
1 & \mbox{if $R \in C$ with probability x$_C$ }\\
0 & \mbox{if $R \in D$ with probability x$_D$ }
\end{array} \right. \]
where $A$, $B$ are the two types of atoms randomly occupying sublattice $I$ and 
$C$, $D$ are those occupying sublattice $II$.

{\par} The diagonal term $\underline{\varepsilon_{R}}$ for such a binary 
distribution  can be written as,
\begin{eqnarray}
\underline{\varepsilon}_{R} &=& 
\left(\begin{array}{cc}
\epsilon_{A}^{I} & t_{AC} \\
t_{AC}           & \epsilon_{C}^{II}  \end{array}\right) n_{R}^{I} n_{R}^{II}
+
\left(\begin{array}{cc}
\epsilon_{A}^{I}& t_{AD} \\
t_{AD}          & \epsilon_{D}^{II}  \end{array}\right) n_{R}^{I} (1-n_{R}^{II})
+\nonumber\\
&&\left(\begin{array}{cc}
\epsilon_{B}^{I}& t_{BC} \\
t_{BC}          & \epsilon_{C}^{II}  \end{array}\right) (1-n_{R}^{I}) n_{R}^{II}+\nonumber\\
&&\left(\begin{array}{cc}
\epsilon_{B}^{I}& t_{BD} \\
t_{BD}          & \epsilon_{D}^{II}  \end{array}\right) (1-n_{R}^{I}) (1-n_{R}^{II})\nonumber\\
&=& E_1 + E_2\ n_{R}^{I} + E_3\ n_{R}^{II} + E_4\ n_{R}^{I} n_{R}^{II},
\label{eq4}
\end{eqnarray}
where
\begin{eqnarray}
E_1 &=& \left(\begin{array}{cc}
\epsilon_{B}^{I} & t_{BD} \\
t_{BD}  &  \varepsilon_{B}^{II}\end{array}\right);\ \ \
E_2 = \left(\begin{array}{cc}
\delta\varepsilon_1 & t^{(1)} \\
t^{(1)}  &  0 \end{array}\right), \nonumber\\
E_3 &=& \left(\begin{array}{cc}
0  & t^{(2)} \\
t^{(2)}  & \delta\varepsilon_2\end{array}\right);\ \ \
E_4 = \left(\begin{array}{cc}
0        & t^{(3)} \\
t^{(3)}  &  0 \end{array}\right)
\label{eq5} 
\end{eqnarray}
with $\delta\varepsilon_1 = \varepsilon^I_A-\varepsilon^{I}_B$;
with $\delta\varepsilon_2 = \varepsilon^{II}_C-\varepsilon^{II}_D$;
 $t^{(1)}=t_{AD}-t_{BD}$, $t^{(2)}=t_{BC}-t_{BD}$ and $t^{(3)}=t_{AC}+t_{BD}-t_{AD}-t_{BC}$.

{\par} Similarly the off-diagonal term $\underline{V}_{RR'}$ in Eq. \ref{eq3} 
can be expressed as (assuming $t'=0$),
\begin{equation}
\underline{V}_{01}=\underline{V}_{02} =V_1 + V_2\ n_{R}^{I} + V_3\ n_{R'}^{II} + V_4\ n_{R}^{I} n_{R'}^{II},
\label{eq6}
\end{equation}
where
\begin{eqnarray}
V_1 &=& \left(\begin{array}{cc}
0       &  0 \\
t_{BD}  &  0\end{array}\right);\ \ \
V_2  = \left(\begin{array}{cc}
0       &  0 \\
t^{(1)}  &  0\end{array}\right),\nonumber\\
V_3 &=& \left(\begin{array}{cc}
0       &  0 \\
t^{(2)}  &  0\end{array}\right);\ \ \
V_4  = \left(\begin{array}{cc}
0       &  0 \\
t^{(3)}  &  0\end{array}\right).
\label{eq7} 
\end{eqnarray}
$\underline{V}_{03}$ ($=\underline{V}_{04}$) are just the transpose of the 
above matrix $\underline{V}_{01}$. Various $t_{\alpha\beta}$'s in the above 
sets of equation are the hopping energies between various atom types ($\alpha
=A,B$ and $\beta = C,D$) at two sublattices $I$ and $II$ respectively. 
 
{\par} Next we proceed to calculate the configuration averaged Green function (
or the Bloch spectral function) in reciprocal space. We shall generalize the augmented
space formalism (ASF) developed earlier in reciprocal space.\cite{kasr} 
The ASF has been described in great detail
earlier.\cite{Mook_book} We shall indicate the main operational results here and
refer the reader to the above monograph for further details. The first step
is to associate with $n_R^I$ and $n_R^{II}$ two operators $N_R^I$ and $N_R^{II}$ such that their spectral density is the probability density of the random variables. For binary random variables, we have :

\[ N_R^I = \left(\begin{array}{cc} x_B & \sqrt{x_Ax_B} \\
                                  \sqrt{x_Ax_B} & x_A
		 \end{array}\right) \]

Finally, according to 
augmented space theorem,\cite{Mookerjee73} the configuration average of any
function of \{$n_R^{I}$,$n_R^{II}$\} can be written as the matrix element,
in {\it configuration} space, of an operator which is the same functional of 
\{$N_R^{I}$,$N_R^{II}$\}. The augmented space Hamiltonian is built up from Eqns.
(\ref{eq4}) and (\ref{eq6}). %We decide how much distortion will be allowed in this problem. 

\bers \widehat{\mathbf {H}} = \sum_R \left\{ E_1 \widehat{I} + E_2 \widetilde{N}^I_R + E_3 \widetilde{N}^{II}_R + 
E_4\widetilde{N}^I_R\otimes\widetilde{N}^{II}_R \right\}\otimes P_R \nonumber\\
  + \sum_R \sum_{R'}\left\{ V_1 \widehat{I} + V_2 \widetilde{N}^I_R + V_3 \widetilde{N}^{II}_{R'} + 
V_4\widetilde{N}^I_R\otimes\widetilde{N}^{II}_{R'} \right\}\otimes T_{RR'}
\label{asr}
\eers 
with 
\begin{equation}
{N}_R^X=x_\alpha\ p_R^{X\uparrow} + x_{\beta}\ p_R^{X,\downarrow} + \sqrt{x_{\alpha}x_{\beta}} (\tau_{R}^{X,\uparrow\downarrow} + \tau_{R}^{X,\downarrow\uparrow}),
\label{eq9}
\end{equation}
($X=I\ \text{or}\ II$)

\begin{figure*}[t]
\centering
\includegraphics[scale=0.7]{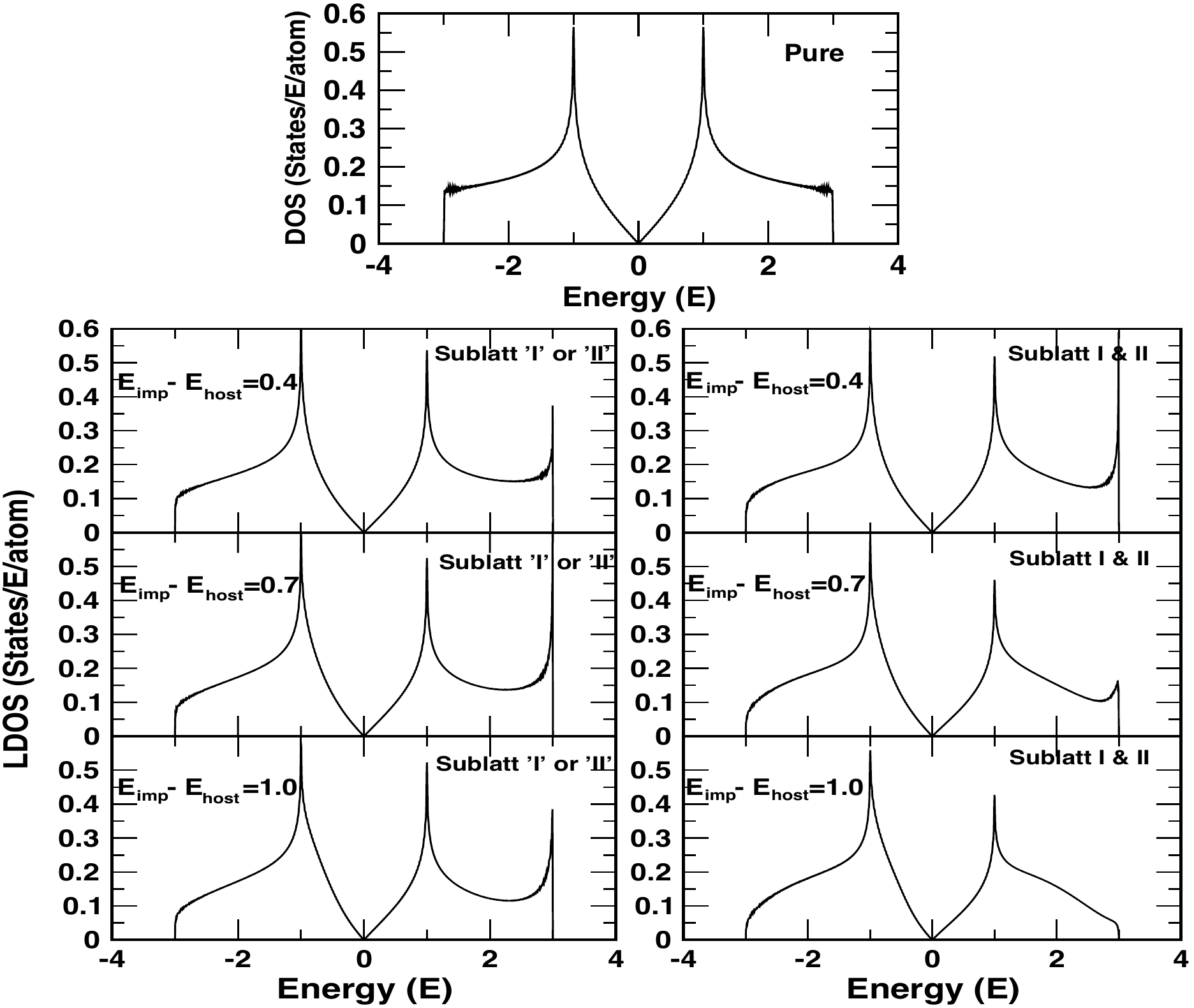}
\caption {Local density of states for pure graphene and graphene with a single impurity. The top panel in the middle is the DOS for pure graphene. Left
 and right panels show the local DOS with single impurity only on sublattice I or II and both I \& II respectively. The panel from top to bottom are the results with increasing 
strength of the impurity potential $\delta{\text{E}}$=$\varepsilon_{\text{imp}}
$-$\varepsilon_{\text{host}}$ }
\label{fig3}
\end{figure*}

\begin{figure}[h]
\centering
\includegraphics[width=6cm]{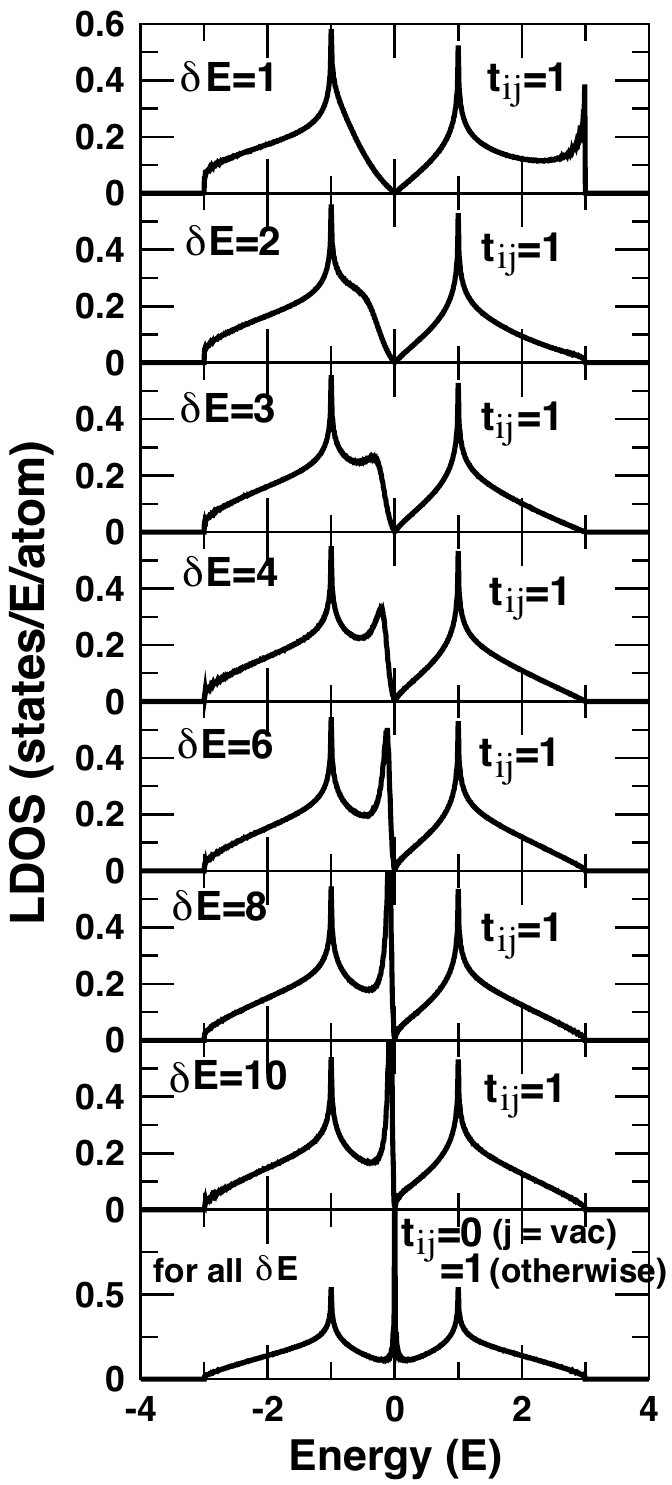}
\caption {Local density of states for graphene with a single vacancy. The vacancy site is modeled by a site with a large repulsive local potential. Technically  we take $\delta{\text{E}} = \varepsilon_{\rm imp}-\varepsilon_{\rm host}$. The figures show consecutive situations with increasing $\delta\text{E}$ as we go from  top to bottom. }
\label{fig4}
\end{figure}

{\par} The configuration averaged Green's function in the reciprocal space is
thus a matrix element of an augmented resolvent given by,

\begin{equation}
\ll G({\bf k},z)\gg  = \langle\{\emptyset\}\otimes{\bf k} \vert (z\widehat{\bf I}-\widehat{\mathbf H})^{-1}\vert\ {\bf k} \otimes \{\emptyset\}\rangle,
\label{eq10}
\end{equation}
 $\vert\ {\bf k} \otimes \{\emptyset\}\rangle$ is an augmented space state in the reciprocal space given by,

\begin{equation}
\vert\ {\bf k} \otimes \{\emptyset\}\rangle = \frac{1}{\sqrt{N}}\sum_R e^{-i{\bf k.R}}\vert\ R \otimes \{\emptyset\}\rangle,
\label{eq11}
\end{equation}
$\vert\ R \otimes \{\emptyset\}\rangle$ is an enlarged basis which is a direct product of the Hilbert space basis \{$R$\} and the configuration space basis $\{\phi_R\}$. The
configuration space $\Phi=\prod_{R}^{\otimes} \phi_R$, takes care of the 
statistical average, is of rank $2^M$  for a system of $M$-lattice sites with
binary distribution.

The recursion follows as a three step generation of a new basis $\{|n>\}$ :

\[|1\rangle = |\k\otimes \{\emptyset\}\rangle\qquad |0\rangle = 0 \]

 \[|n+1\rangle = \widehat{H}|n\rangle - \alpha_n|n\rangle -\beta^2_{n-1}|n-1\rangle\]
\[
\alpha_n(\k) = \frac{\langle n|\widehat{H}|n\rangle}{\langle n|n\rangle}\ \mbox{and }\ \beta_n^2(\bf k) =\frac{\langle n|n\rangle}{\langle n-1|n-1\rangle}\]

The ASR  gives the configuration averaged spectral function as a continued  fraction :

\begin{widetext}
\ber
\ll G({\bf k},z)\gg &=& \frac{1}{\displaystyle z-\alpha_1({\bf k}) -
\frac{\beta_1^2({\bf k})}{\displaystyle z-\alpha_2({\bf k})-\frac{\beta_1^2({\bf k})}{z-\alpha_3({\bf k})-\frac{\ddots}{\displaystyle T(z,{\bf k})}}}}
 =  \frac{1}{z-E_0(\k)-\Sigma(z,\k)}
\eer
\end{widetext}

$T(z,\k)$ is a continued fraction {\it terminator} as proposed by Beer and Pettifor.\cite{BF} The spectral function peaks are 
decided by $\Re e\{\Sigma(E,\k)\}$ and the imaginary part of $\Sigma$ gives the width related to the disorder induced lifetimes.

{\par} The configuration averaged Bloch spectral function is  given by,
\begin{equation}
\ll A({\bf k}, E)\gg = -\frac{1}{\pi} \lim_{\delta\rightarrow 0^{+}}  \Im m \{ \ll G({\bf k}, E+i\delta) \gg\}
\label{eq12}
\end{equation}

{\par} The configuration averaged density of states (DOS) is,
\begin{equation}
\ll n(E)\gg = \frac{1}{\Omega_{BZ}} \int_{BZ} d{\bf k}\ \ll A({\bf k}, E) \gg
\label{eq13}
\end{equation}
{\par} The electronic dispersion curves are obtained by numerically calculating
the peak E-position of the spectral function. The full-widths at half maxima 
(FWHM) are also calculated from the disorder broadened Bloch spectral function.

\section{Results and Discussion}

In the following subsections, we shall present our results for  graphene with impurities,
vacancies, diagonal disorder alone, and with the simultaneous presence of diagonal and off-diagonal disorder. 
The effects of various strengths of impurity potentials  on two inequivalent 
sublattices will be shown via changes in the shape of the DOS. The changes in 
the topology of Dirac-cone dispersion, disorder-induced  FWHM and the DOS will be shown for various strengths of diagonal disorder. In the most general 
case of diagonal and  off-diagonal disorder, we consider three interesting limiting
cases: (i) strong diagonal and weak off-diagonal disorder (ii) strong
off-diagonal and weak diagonal disorder and (iii) strong diagonal as well as 
off-diagonal disorder. The interesting interplay between these different kinds
of disorder in graphene reveals a discontinuous type of band near the 
$\Gamma$-point in the third limiting case.

\begin{figure*}[t]
\centering
\includegraphics[scale=0.9]{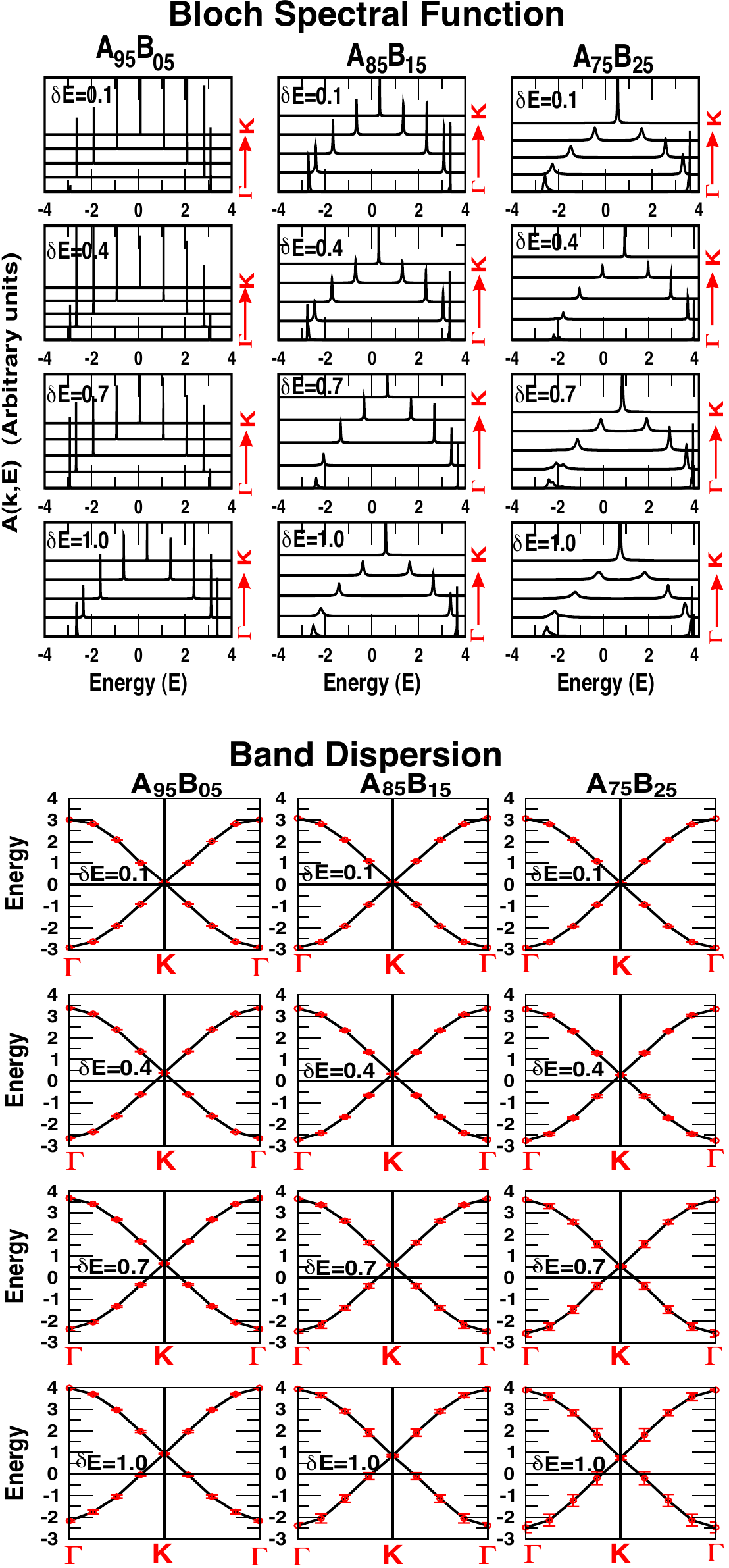}
\caption {(Color online) The averaged spectral functions (upper set of panels) and the complex bands (lower set of panels) near the Dirac point. These are all
for pure diagonal disorder at three different alloy compositions (left to right) and four disorder strengths $\delta\text{E}$ (top to bottom). The 
(red) error bars show how the disorder induced lifetimes vary across the 
samples.}
\label{fig5}
\end{figure*}

\begin{figure*}[t]
\centering
\includegraphics[scale=0.65]{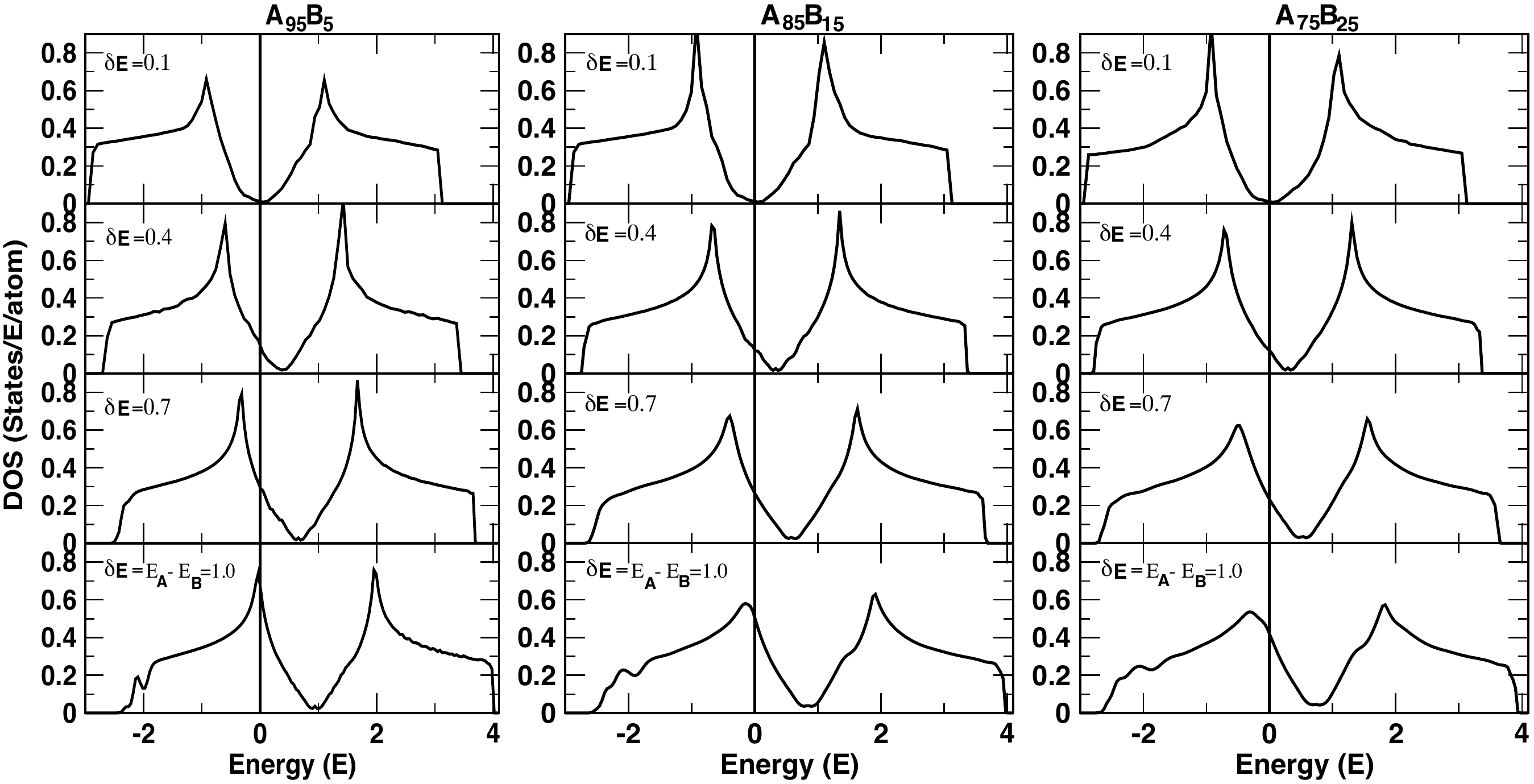}
\caption { Total DOS for the same set of disorder strengths
 $\delta{\text{E}}$ for three alloys A$_{x_A}$B$_{x_B}$ as in Fig. \ref{fig5}.
Due to homogeneous diagonal disorder on both the sublattices $I$ and $II$, the
individual projected DOS on them are same in this case. }
\label{fig6}
\end{figure*}

% Effects of diagonal disorder on electronic dispersion......... 
%\begin{figure*}[t]
%:\%centering

\subsection{Impurities in Graphene}

% Impurity effects on DOS......... 
In Fig. \ref{fig3}, we display the DOS with different strengths of the single 
impurity potentials on different inequivalent sublattices. The top figure in the 
middle panel is the DOS for pure graphene. The left and  right 
panels show local DOS with a single impurity put on the sublattice $I$ or $II$
and both respectively. The strength of the impurity potential (relative to the
host lattice) increases from top to bottom panels (i.e. $\delta{\text{E}}$ =  $\varepsilon_{\text{imp}}$ - $\varepsilon_{\text{host}}$ = $0.4$, $0.7$ and $1.0$). All these 
calculations are
done with a fixed hopping parameter t=1. We notice  changes in 
the shape of the hump  and the van-Hove singularities as 
the strength of the impurity potential increases. Although the effects are
small, but are clearly visible for the case of $\delta{\text{E}}$=1.0, where
 the local environment around the impurity site feels the strongest scattering.
With the introduction of the impurity, the symmetry of the DOS around the Dirac point is lost. At these impurity levels,   both the left and right panels show
the formation of an impurity peak near the upper band edge. With increasing 
disorder this impurity peak moves into the band and disappears. Again, 
at these strengths there is no perceptible changes to the linear structure of the Dirac point.  Similar results have been obtained previously for such models of impurities. This provides the correctness of our new formulation.

\subsection{Single vacancies in graphene}

An extension of the single impurity is the single vacancy in the model. The
vacancy is modeled by a site with a large repulsive local potential. 
This is shown in Fig. \ref{fig4}. Notice that, with
increasing $\delta \text{E}$, the rightmost impurity peak at around the top band-edge moves into
the band. Most of the changes occur around the Dirac point at E=0. At  $\delta \text{E} =2 > t(=1)$ the symmetry of the Dirac point
gets broken with the appearance of another  mild peak. The origin of this `zero mode' peak has been extensively discussed by Pereira {\emph et al.}\cite{Pereira} in detail. This peak grows with
increasing disorder  until at $\delta \text{E} \simeq 30$, where it takes the form of a delta
function like structure at the Dirac point. The bottom panel shows the result
for an ideal vacancy, where $t_{ij}=0$ for $t$ connecting the vacancy to the graphene lattice, or in other words a completely inaccessible `hard' vacancy. The symmetry of the
Dirac point is restored and the vacancy peak sits exactly at the Dirac point.
This is exactly the same behavior reported by Pereira {\emph et al.}\cite{Pereira} who used either the CPA or direct real space recursion.

\subsection{Diagonal disorder}

\begin{figure*}[t]
\centering
\includegraphics[scale=0.9]{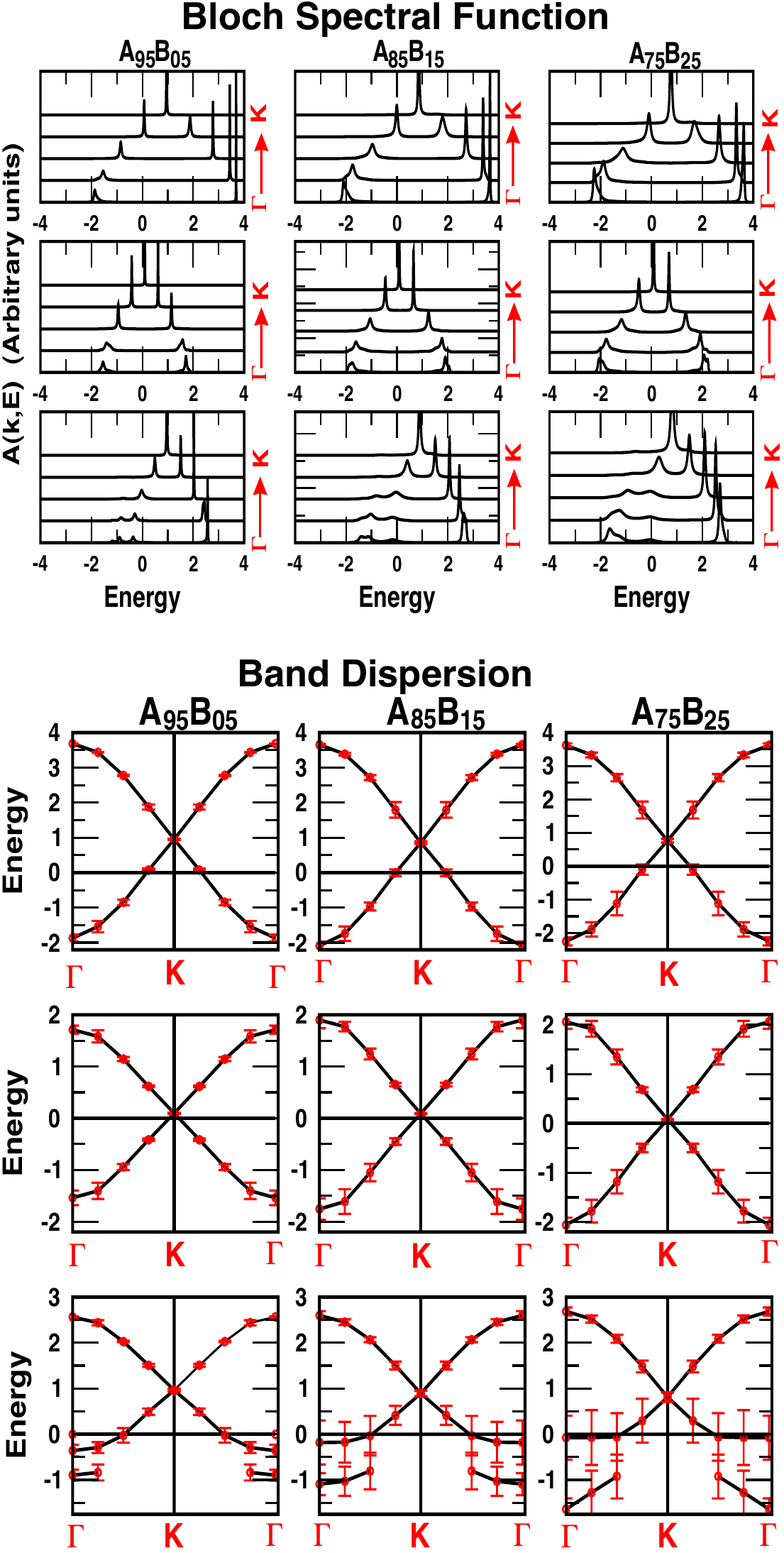}
\caption {(Color online) Same as Fig. \ref{fig5}, but with the inclusion of both diagonal and off-diagonal disorder.  The three panels for each alloy indicate the results with coupled diagonal and off diagonal disorders as described in the text.}
\label{fig7}
\end{figure*}

\begin{figure*}[t]
\centering
\includegraphics[scale=0.7]{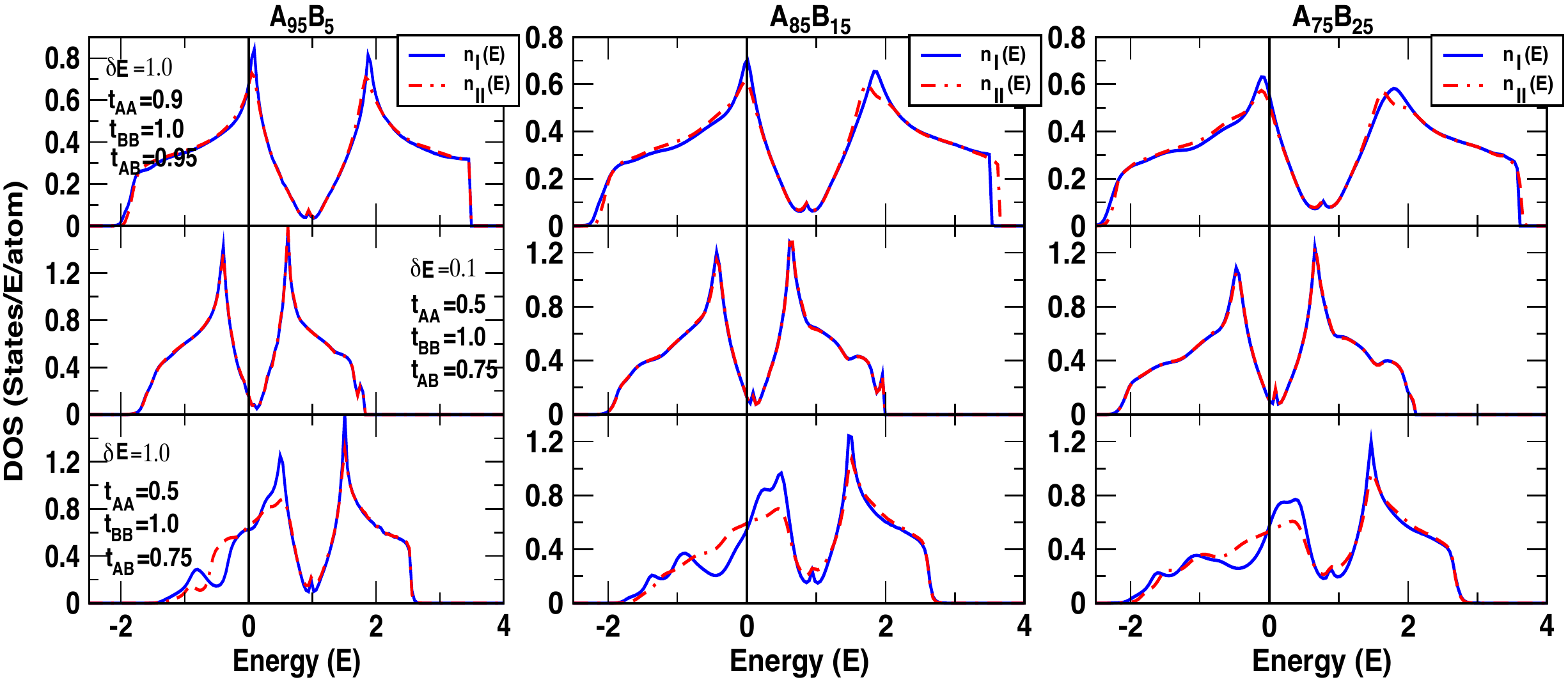}
\caption {(Color online) Sublattice projected DOS for the same  three set of diagonal+ off-diagonal disorder strengths for three alloys A$_{x_A}$B$_{x_B}$ as in Fig. \ref{fig7}. The projected DOS on the two sublattices n$_I$ and n$_{II}$, in this case, are different due to the obvious reason arising from different random t$_{ij}$  interactions. }
\label{fig8}
\end{figure*}

First we shall take up purely diagonal disorder problems : those problems which
can be taken up by earlier suggested methodologies. Of course, our augmented 
space recursion in reciprocal space give us additional information about 
the disorder induced life-times of the Bloch states. 
 In Fig. \ref{fig5}, we display the configuration averaged Bloch spectral
function (upper panels) (given by Eq. (\ref{eq12})) and the corresponding dispersion Energy vs. $k$ (lower panels) along
$\Gamma-\text{K}-\Gamma$ symmetry line for three different alloys A$_{x_{A}}$B$_{x_B}$ ($x_B=5\%, 15\%$ and $25\%$) with various diagonal disorder strengths.
 The two sublattices $I$ and $II$ are homogeneously disordered, such that 
$x_A=x_C$, $x_B=x_D$ and $\varepsilon_A^{I}=\varepsilon_C^{II}$, $\varepsilon_B^{I}=\varepsilon_D^{II}$. For each alloy case, the panels from top to bottom indicate the 
results with  increasing strength of diagonal disorder (i.e. $\delta\text{E}$=
$\varepsilon_A - \varepsilon_B$ = $0.1,0.4,0.7$ and $1.0$). The hopping integral $t$ is 
chosen to be 1 here, so there is no off-diagonal disorder. For each disorder strength $\delta\text{E}$ in a particular 
alloy,  the upper panels show the averaged Bloch spectral function 
 at five {\bf k}-points along the
high symmetry direction $\Gamma-$K. The corresponding Dirac dispersions
are shown in the lower set of  panels along $\Gamma-\text{K}-\Gamma$ line. The first
thing to note is that the spectral function modifies quickly from sharp 
near $\delta$-functions to Lorentzian shapes with increasing disorder strength
$\delta{\text{E}}$  as well as increasing alloy concentration
$x_B$. In addition, the function gets more and more
asymmetric with increasing $\delta{\text{E}}$. Such asymmetries can be described as a
tendency of more scattering to occur near the resonance energies around 
$\Gamma$. In other words, line shapes around $\Gamma$ tend to have a weak
second peak or wide tail over the resonance region. For the present diagonal 
disordered case, the Dirac point is simply shifted by an average energy
$\langle\epsilon\rangle=x_A \epsilon_{A}^{I} + x_B \epsilon_{B}^{II}$

% Effects of diagonal disorder on DOS......... 
{\par} The corresponding total DOS for the same set of disorder
strengths ($\delta\text{E}$) and the alloy concentrations ($x$) are shown in Fig. \ref{fig6}. The individual projected DOS on the two sublattices $I$ and $II$
in this case are same, because we have maintained uniform diagonal disorder
on both the sublattices. However, the present theory is equally capable of 
treating the two sublattices differently with a different nature of disorder
on them. In that case, the two inequivalent sublattices will have different
projected quantities. Looking at Fig. \ref{fig6}, one can notice an exactly 
similar shift of the Dirac point (to the average $\langle\epsilon\rangle$) in
the DOS as shown in the dispersion. The disorder effects are pronounced
around the Dirac-point energy $\langle\epsilon\rangle$ and get milder around 
the hump below $\delta\text{E}=0.7$. Above this disorder strength, the left
band edge starts to show up extra features with a dip at around $\text{E}=-2$,
(as shown in the bottom panel for all the three alloy concentration). The results are qualitatively similar to the CPA works done earlier\cite{Pereira} but differ in  quantitative details. 

\subsection{Off-diagonal disorder}

% Effects of off-diagonal disorder on electronic dispersion......... 

We now turn to the cases with off-diagonal disorder. Such problems cannot be dealt
with within the CPA. Also, direct calculation of the averaged spectral functions and disorder induced lifetimes is also not feasible with other techniques and the strength of the ASR comes to the fore. In addition we should note that in our model, diagonal and off-diagonal disorders are correlated : e.g. if the atom A occupies the site $i$ with probability $x_A$ and atom B occupies the site 
$j$ with probability $x_B$, then $t_{ij}$ {\it has to be} $t_{AB}$ with probability 1.
 Although the present theory is equally capable of 
investigating other interesting cases (e.g. inhomogeneous disorder, 
pseudo-binary type disorder etc.), here we have chosen to explore three 
 cases which should reflect the behavior of a variety of the 
realistic materials. The three cases are:
\begin{itemize}
\item {Strong diagonal and weak off-diagonal disorder; with parameters $\delta
{\text{E}}=\epsilon_{A}^{I}-\epsilon_{B}^{I}=\epsilon_{C}^{II}-
\epsilon_{D}^{II}=1.0$, $t_{AC}=1.0,t_{BD}=0.9$ and $t_{AD}=t_{BC}=0.95$. }
\item {Weak diagonal and strong off-diagonal disorder; with parameters $\delta
{\text{E}}=0.1$, $t_{AC}=1.0,t_{BD}=0.5$ and $t_{AD}=t_{BC}=0.75$. }
\item {Strong diagonal as well as strong off-diagonal disorder; with parameters
 $\delta{\text{E}}=1.0$, $t_{AC}=1.0,t_{BD}=0.5$ and $t_{AD}=t_{BC}=0.75$. }
\end{itemize}

{\par} The results for these three cases are shown in the top, middle and bottom
 panels of Fig. \ref{fig7} respectively, for the same three alloys A$_{x_A}$B$_{x_B}$ as 
before. Other details are same as in Fig. \ref{fig5}. Notice that unlike the
diagonal disordered case, effects of both diagonal and off-diagonal disorder 
are much more dramatic. In addition to highly asymmetric nature, the Bloch
spectral function is found to have a  double peaked structure in the extreme
 case of strong diagonal and off-diagonal disorder (shown in the bottom panels).
 Such doubly peaked line shape introduces extra discontinuous bands in the 
dispersion curve. Such a structure had been seen before in phonon problems\cite{phonon} which
also have intrinsic off-diagonal disorder in the dynamical matrices. There it arose because of
resonant modes. Here too we shall give a similar explanation.
These dispersion at resonance have  relatively large FWHM's 
 and it will be interesting to choose a realistic material of similar disorder properties and investigate the 
experimental outcome. 
% Effects of diagonal+off-diagonal disorder on DOS......... 

{\par}Figure \ref{fig8} shows the sublattice projected DOS for the same three limiting cases for the three alloys as above. The solid blue and the dashed red
lines indicate the projected DOS on the sublattices $I$ and $II$ respectively.
Because of the random hopping (off-diagonal) interaction in this case, the two
sublattices acquire different environment around it, and hence possess different
projected quantities on them. As expected, the DOS in these cases have  large smearing. The effective environment around the two sublattices is maximally different from each other in the extreme case of both strong diagonal+off-diagonal disorders (as shown in the bottom panels), as evident from the large difference between their projected DOS. Interestingly, for this particular case, the consequence of the discontinuous bands in the -ve energy range (bottom panels of Fig. \ref{fig7}) shows up via a dip in the DOS along with a much larger smearing.   
Apart from this extreme case, the  Dirac point for all the other cases  
has moved in exact
accordance with that of the band shift as in Fig. \ref{fig7}. The topology of
the DOS on the two sides of the Dirac point are very different from each other 
specially in the case of strong diagonal and off-diagonal disorder (bottom
panels). In totality, the effects of off-diagonal disorder is very different from that of diagonal disorder (as a comparison between Figs. \ref{fig6} 
and \ref{fig8} will show). Treatments of off-diagonal disorder
is straight-forward and accurate in the ASR formalism. 
\vskip 1cm
\section{Conclusion}
We present a theoretical model to study the effects of diagonal
and off-diagonal disorder in graphene on an equal footing. To our knowledge this is the first 
theoretical framework to reliably take into account the effects of off-diagonal
disorder in describing the spectral properties of graphene. We show how the topology of the Dirac
dispersion, and the location of Dirac point change with the strength of
disorder and impurity concentration. Interestingly, the dispersion in case of
strong diagonal and off-diagonal disorder tends to have an extra discontinuous
band which is rather uncommon in the graphene fermiology with simple disorder.
As such we propose to verify such effects in the electronic dispersion by 
setting up an experiment on a similar realistic graphene system, where both 
the diagonal and the off-diagonal disorder are strong. We believe that such a 
study may provide a deeper insight into the physics and materials perspective
of graphene. Finally, we want to state that our formulation is quite general and can be applied 
to the case of other 2D materials, e.g., BN in presence of disorder. 
\vskip 1cm

%%%%%%%%%%%%%%%%%%%%%%%%%%%%%%%%%%%%%%%%%%%%%%%%%
%                     Acknowledgements
%%%%%%%%%%%%%%%%%%%%%%%%%%%%%%%%%%%%%%%%%%%%%%%%%
\emph{Acknowledgments:} AA acknowledges support from the U.S. Department of Energy BES/Materials Science and Engineering Division from contracts DEFG02-03ER46026 and Ames Laboratory (DE-AC02-07CH11358), operated by Iowa State University. This work was done under the HYDRA collaboration between the research groups..
%%%%%%%%%%%%%%%%%%%%%%%%%%%%%%%%%%%%%%%%%%%%%%%%%%
%                     Reference Page
%%%%%%%%%%%%%%%%%%%%%%%%%%%%%%%%%%%%%%%%%%%%%%%%%%
\vspace{-0.5cm}
%\bibliography{Aftab}

\end{document}